# Approaching High-efficiency Spatial Light Modulation with Lossy Phase-change Material

LUOYAO CHU,[1] YAN LI,[2,4] SHUNYU YAO,[1] YURU LI,[1] SIQING ZENG[1] AND ZHAOHUI LI[1,3,5]

[1]*Key Laboratory of Optoelectronic Materials and Technologies, School of Electrical and Information Technology, Sun Yat-sen University, Guangzhou 510275, China*
[2]*School of Microelectronics Science and Technology, Sun Yat-sen University, Zhuhai, 519000, China*
[3]*Southern Marine Science and Engineering Guangdong Laboratory (Zhuhai), Zhuhai 519000, China*
[4]*e-mail: liyan329@mail.sysu.edu.cn*
[5]*e-mail: lzhh88@mail.sysu.edu.cn*

**Abstract:**
The prevalent high intrinsic absorption in the crystalline state of phase-change materials (PCMs), typically leads to a decline in modulation efficiency for phase-change metasurfaces, underutilizing their potential for quasi-continuous phase-state tuning. This research introduces a concise design approach that maximizes the exploitation of the quasi-continuous phase transition properties of PCM, achieving high-efficiency spatial light modulation. By optimizing the metasurface design, the phase modulation process is strategically localized in a low crystallization ratio state, significantly reducing the impact of material absorption on modulation efficiency. Utilizing GSST as an example, numerical simulations demonstrate a minimum reflectance of 46.5% at the target wavelength of 1550 nm, with a phase modulation depth of 246.6°. The design is fruther extended to dynamic polarization modulation by incorporating structural anisotropy, enabling independent control of amplitude and phase modulation in orthogonal polarization directions. This strategy not only circumvents the efficiency limitations of crystalline states but also harnesses the full potential of PCMs for multifunctional photonic applications.

## 1. Introduction

As a crucial technology for applications in optical communications, optoelectronic displays, and LiDAR, multi-dimensional modulation of spatial light fields remains a core topic in modern optical research [1-3]. Metasurfaces, two-dimensional planar devices derived from three-dimensional metamaterials, have emerged as a research hotspot in this field due to their compact size, ease of integration, and high tunability [4-6]. Recently, active metasurfaces have been developed by integrating various active materials, paving the way for low-cost, lightweight, and integrated spatial light modulators (SLMs) [7, 8]. Commonly used active materials include transparent conductive oxides [9-11], liquid crystals [12-14], two-dimensional materials [15-17], semiconductors [18-20], vanadium dioxide [21-23] and chalcogenide phase-change materials (PCMs) [24-26]. Among these materials, chalcogenide PCMs are particularly noteworthy for their ability to reversibly transform between crystalline and amorphous states. Chalcogenide PCMs exhibit several advantages, including non-volatility [27, 28], rapid phase-change rates [29], low power consumption [30], long cycle life [31, 32], compatibility with CMOS processes [33] and significant modulation of the refractive index [34]. Consequently, they hold great promise for applications in the design and manufacture of high-performance SLMs.

To date, numerous studies have focused on the modulation of spatial light fields using phase-change metasurfaces. Early research primarily employed the classic $Ge_2Sb_2Te_5$ (GST) as the active material. As a well-established PCM, GST has been widely employed in commercial storage media [35, 36]. However, GST exhibits a relatively high extinction coefficient in the visible to near-infrared range [34]. In active metasurfaces, the excitation of resonant modes within the microstructure further enhances the intrinsic absorption of the active material, resulting in strong coupling between amplitude and phase modulation [37-39]. Recently, the emergence of lossless chalcogenide PCMs, such as $Sb_2S_3$ (SbS) and $Sb_2Se_3$ (SbSe), has garnered increasing attention for their potential in high-efficiency spatial light field modulation using phase-change metasurfaces. These materials can maintain a near-zero extinction coefficient in the short-wave infrared to near-infrared range during phase transitions. Theoretically, optimizing the metasurface design can effectively mitigate intrinsic material absorption and structural losses, thereby enabling high-efficiency phase modulation of the spatial light field, potentially approaching 100% [40]. However, previous studies have revealed that SbS and SbSe suffer from random lattice distributions during phase transitions, resulting in uneven refractive index distributions. This results in significant lattice scattering, greatly constraining the modulation efficiency of the metasurface [41-43]. In addition, a class of low-loss chalcogenide PCMs is represented by $Ge_2Sb_2Se_4Te_1$ (GSST) and $Ge_2Sb_2Se_5$ (GSS). These materials are synthesized by doping GST with the more electronegative selenium (Se), and they are considered potential replacements for GST. Although their amorphous states exhibit low-loss properties, a relatively large extinction coefficient remains in their crystalline state [44]. Consequently, current research emphasizes the application of these materials in switch-type active integrated photonic devices to fully exploit their excellent performance in the amorphous state [24, 45]. However, this approach sacrifices the quasi-continuous tunability of the phase states of chalcogenide PCMs. Thus, reducing the impact of intrinsic absorption introduced after crystallization on metasurface devices to achieve efficient quasi-continuous spatial light field modulation has become crucial for advancing practical applications. Furthermore, low-loss properties in the amorphous state and high-loss characteristics in the crystalline state represent common challenges encountered by chalcogenide PCMs. Therefore, effectively reducing or avoiding the impact of chalcogenide PCMs on metasurface devices under high crystallization ratios is essential for achieving efficient spatial light field modulation based on phase-change metasurfaces.

In this work, a simple design concept for phase-change metasurfaces is proposed to reduce the impact of intrinsic absorption of chalcogenide PCMs on the efficiency of spatial light field manipulation under high crystallization ratios. The design optimizes the metasurface such that the resonance center wavelength of the dielectric resonance mode in the amorphous state is close to the target wavelength. This ensures a certain phase modulation depth while improving the minimum reflectance of the metasurface at the target wavelength during the phase transition, thereby enhancing the efficiency of spatial light field modulation. A reflective metasurface based on GSST was designed and utilized as an example to verify the design concept. Phase modulation of the reflected light field is achieved by exciting the electric dipole (ED)-dominated dielectric resonance mode in the GSST microstructure while ensuring that it remains overcoupled during the phase transition. Simultaneously, by varying the lateral dimensions of the GSST microstructure, the resonance center wavelength of the dielectric resonance mode in the amorphous state is adjusted to be close to the target wavelength. The optimized metasurface achieved a minimum reflectance of 46.5% at the target wavelength of 1550 nm while covering a phase modulation depth of 246.6°. Furthermore, based on the gradient-phase metasurface, a +1st order diffraction efficiency of 51.3% was demonstrated at a deflection angle of 4.29°. The anisotropy of artificial microstructures was also considered, with selective introduction of amplitude-dominated and phase-dominated quasi-continuous modulation of phase change, further extending the current design concept to the field of dynamic polarization modulation. Notably, the proposed design concept of phase-change metasurfaces addresses the common challenges associated with chalcogenide PCMs and opens new avenues for achieving high-efficiency spatial light field modulation.

## 2. Principle

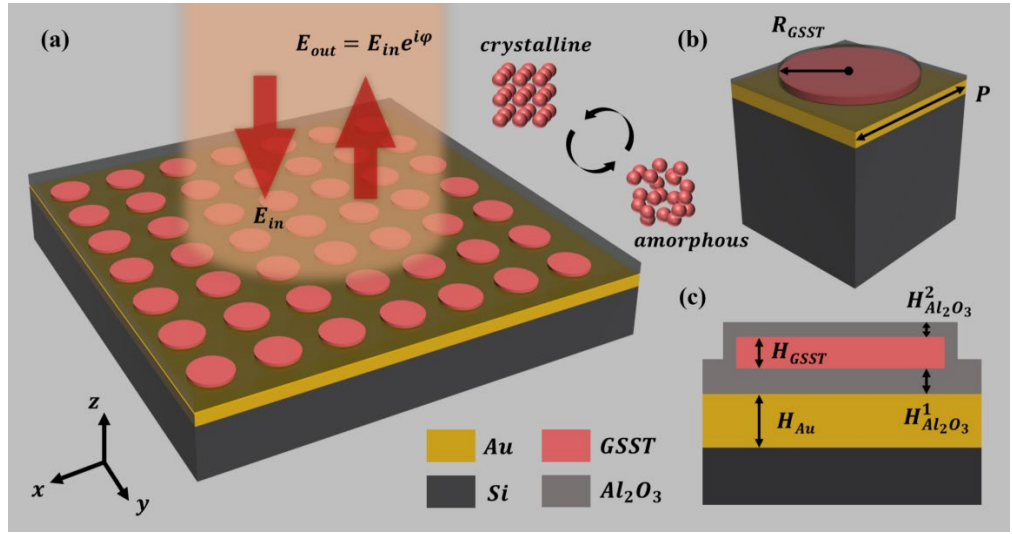

Fig. 1. Schematic diagram illustrating the structure of the reflective metasurface based on GSST.
(a) Three-dimensional structure. (b) Three-dimensional representation of the metasurface unit.
(c) Longitudinal cross-section of the metasurface unit.

In this research, a reflective metasurface based on GSST is designed and proposed using the classic metal-insulator-dielectric (MID) structure. The unit structure of this metasurface, from bottom to top, consists of a Si substrate, an Au reflective layer, an $Al_2O_3$ isolation layer, a GSST cylinder, and an $Al_2O_3$ protective layer, as illustrated in Fig. 1. When the spatial light field is incident, specific dielectric resonance modes are excited within the GSST cylinder. The mirroring effect of the Au reflective layer enables these resonance modes to couple with the mirror resonance modes through near-field electromagnetic interactions, thereby enhancing control over the spatial light field. By optimizing the geometric parameters of the metasurface unit, it is ensured that the dielectric resonance modes supported by the GSST cylinder during the phase transition remain overcoupled, thereby effectively modulating the phase of the reflected light field. However, a common challenge encountered by current active metasurfaces is that the dielectric resonance modes supported by the microstructures are typically localized within the near field of the active materials, rendering them highly sensitive to the intrinsic absorption of these materials and consequently limiting the efficiency of spatial light field modulation. Although GSST exhibits favorable refractive index modulation capabilities, it still confronts the common challenges associated with chalcogenide PCMs: low loss in the amorphous state and high loss in the crystalline state, as illustrated in Fig. 2(a) and 2(b). Consequently, researchers have concentrated on utilizing GSST to develop switch-type active photonic devices that leverage its low-loss characteristics in the amorphous state to achieve efficient switching functionalities. Notably, although the intrinsic absorption of GSST in the crystalline state limits the modulation efficiency of the spatial light field, its excellent quasi-continuous tunability in the phase state presents an opportunity to mitigate the impact of the PCMs on the modulation efficiency of the spatial light field within the metasurface under high crystallization ratios. In general, the Lorentz-Lorenz equation effectively describes the quasi-continuous variation of the equivalent complex refractive index of the PCMs during the phase transition [46, 47]. Based on this, this paper presents the relationship between the refractive index, extinction coefficient, and crystallization ratio at a wavelength of 1550 nm, as illustrated in Fig. 2(c). This indicates that as the crystallization ratio increases, both the refractive index and extinction coefficient gradually rise, peaking in the fully crystalline state. Therefore, this research introduces a straightforward design concept for metasurfaces that aims to fully exploit the gradual variation of the complex refractive index of GSST during the phase transition, thereby minimizing the impact of intrinsic absorption at high crystallization ratios on the efficiency of spatial light field modulation.

To elucidate the proposed design concept of the metasurface, we present the theoretically typical reflectance and phase spectra of the GSST-based reflective metasurface in both crystalline and amorphous states, as illustrated in Fig. 2(d). The metasurface demonstrates substantial phase modulation of the reflected light field across a bandwidth that encompasses the resonance center wavelength of the dielectric resonance mode. By comparing the modulation effects at the two target wavelengths at the extremes of the bandwidth range, we observed that while both wavelengths experience the same phase modulation depth, the amplitude change at $\lambda_2$ is considerably greater than at $\lambda_1$. This phenomenon can be attributed to the proximity of $\lambda_1$ to the resonance center wavelength of the mode supported by the amorphous phase-change microstructure. The critical region for phase modulation occurs in a low crystallization ratio state characterized by a small extinction coefficient, leading to reduced absorption at $\lambda_1$, as illustrated in Fig. 2(e). Conversely, $\lambda_2$ is near the resonance center wavelength of the mode supported by the phase-change microstructure in the crystalline state, and the critical region for phase modulation occurs in a high crystallization ratio state characterized by a large extinction coefficient. This results in significant absorption at $\lambda_2$ and leads to strong coupling between amplitude and phase modulation, as illustrated in Fig. 2(f). Therefore, this design concept not only preserves sufficient phase modulation depth during the phase transition but also significantly enhances the minimum reflectance of the metasurface at the target wavelength throughout the phase transition of the phase-change microstructure. This improvement increases the efficiency of the metasurface in regulating the spatial light field.

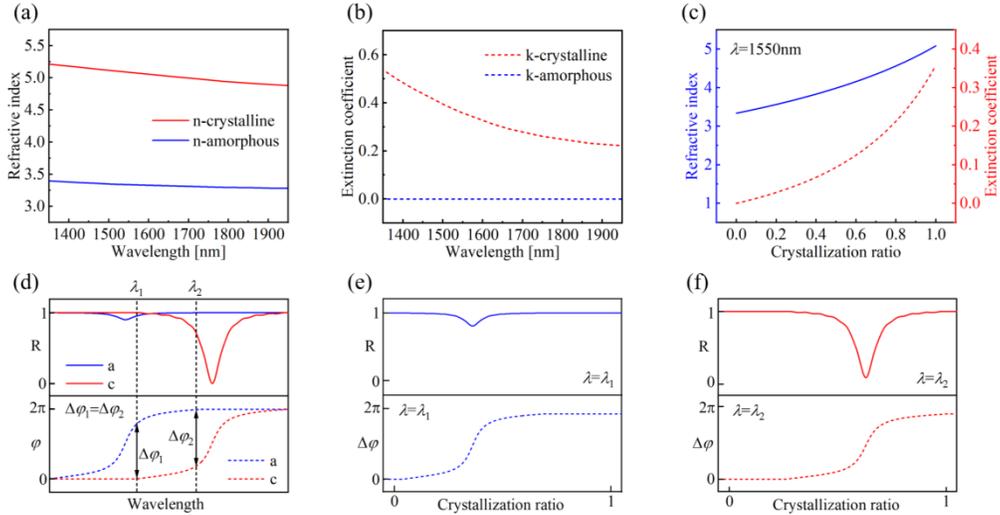

Fig. 2. (a) Refractive indices of GSST in the amorphous and crystalline states. (b) Extinction coefficients of GSST in the amorphous and crystalline states. (c) Refractive indices and extinction coefficients of GSST as functions of the crystallization ratio at a wavelength of 1550 nm. (d) Theoretically typical reflectance and phase spectra of a reflective metasurface based on GSST (crystalline and amorphous states). (e) Reflectance and phase modulation as functions of the crystallization ratio at $\lambda_1$. (f) Reflectance and phase modulation as functions of the crystallization ratio at $\lambda_2$.

## 3. Results

Based on the proposed metasurface design concept, the structural parameters of the metasurface unit were adjusted to ensure that the incident light field effectively excites the electric dipole (ED)-dominated dielectric resonance mode in the GSST cylinder, as detailed in Supplementary Materials S1 and S2. On this basis, the optical response of the metasurface before and after optimization was systematically compared and analyzed by adjusting the radius of the GSST cylinder, for example. The radii of the GSST cylinders before and after optimization were 190 nm and 245 nm, respectively, while the thicknesses of the GSST cylinders, Au reflective layer, $Al_2O_3$ isolation layer, and $Al_2O_3$ protective layer were fixed at 100 nm, 100 nm, 50 nm, and 20 nm, respectively. Additionally, the period of the metasurface unit structure was fixed at 1150 nm. For the metasurface before optimization, the resonance center wavelength of the dielectric resonance mode was shifted from 1388 nm to 1596 nm during the phase transition, resulting in phase modulation of the reflected light field within this wavelength range, as shown in Fig. 3(a) and 3(b). However, the target wavelength of 1550 nm is close to the resonance center wavelength in the crystalline state, which causes the core region of phase modulation to occur in a high crystallization ratio state with a large extinction coefficient, leading to increased absorption at the target wavelength. Consequently, the minimum reflectance of the metasurface at 1550 nm during the phase transition was only 1.1%, significantly limiting the efficiency of spatial light field modulation, as shown in Fig. 3(c). The strong coupling of amplitude and phase modulation imposes significant limitations on phase selection. To achieve efficient quasi-continuous phase modulation, high crystallization ratio regions must be avoided. However, this strategy would significantly sacrifice the phase modulation depth. Nevertheless, this metasurface demonstrates good amplitude quasi-continuous modulation characteristics, making it suitable for use as a switching-type active photonic device. After optimization, the resonance center wavelength of the dielectric resonance mode supported by the GSST cylinder in the amorphous state was adjusted to 1506 nm, bringing it closer to the target wavelength of 1550 nm. At this point, the core region of phase modulation is situated in a low crystallization ratio state with a small extinction coefficient, significantly reducing absorption at the target wavelength, as shown in Fig. 3(d) and 3(e). In this scenario, the minimum reflectance of the metasurface at 1550 nm during the phase transition was significantly enhanced to 46.5%, substantially improving the efficiency of spatial light field modulation at this wavelength and allowing for more flexible phase selection, thereby achieving efficient quasi-continuous phase modulation. Furthermore, the optimized design ensures a phase modulation depth of 246.6°, enabling the metasurface to control the distribution of the spatial light field more accurately and flexibly, as shown in Fig. 3(f). It is noteworthy that this metasurface design concept primarily mitigates the influence of intrinsic absorption from the chalcogenide PCM in the high crystallization ratio state on the efficiency of spatial light field modulation, whereas the Au reflective layer is not the main source of energy absorption. The changes in its absorption before and after optimization are minimal, constituting a small proportion of the total absorption, as detailed in Supplementary Materials S3 and S4. In summary, designing the resonance center wavelength of the dielectric resonance mode of the phase-change microstructures in the amorphous state to be close to the target wavelength ensures that the core region of phase modulation occurs in a low crystallization ratio state where the extinction coefficient is small. The key to this strategy is concentrating phase modulation in regions of low absorption, which not only ensures sufficient phase modulation depth but also significantly reduces the influence of intrinsic absorption in high crystallization ratio states on the efficiency of spatial light field modulation. Furthermore, by fully utilizing the quasi-continuous phase transition properties of chalcogenide PCMs, energy utilization can be effectively balanced to reduce unnecessary energy loss, significantly enhancing the spatial light field modulation efficiency of the metasurface.

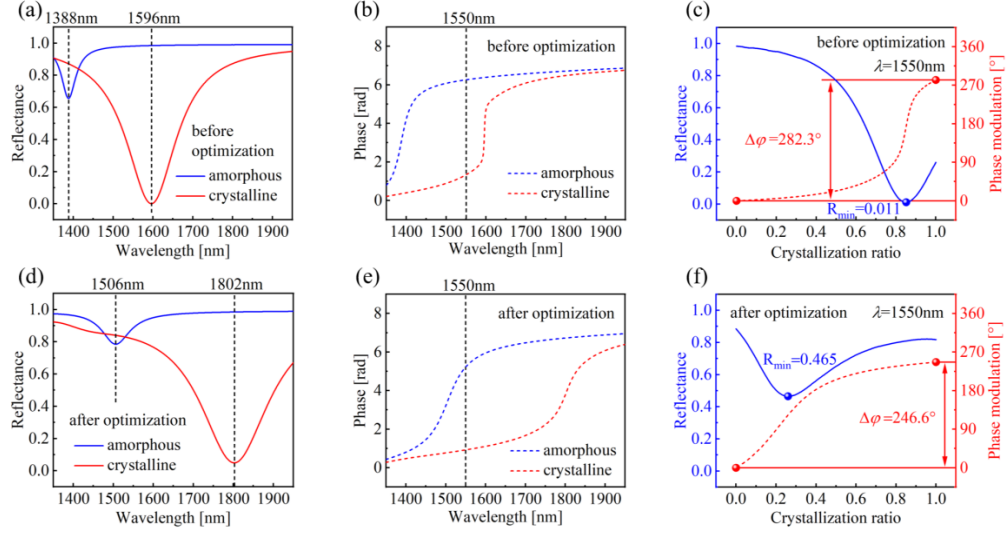

Fig. 3. Comparison of the optical response of the phase-change metasurface before and after optimization. (a) Reflectance spectrum and (b) phase spectrum of the phase-change metasurface before optimization (amorphous and crystalline states). (c) Reflectance and phase modulation as functions of crystallization ratio at 1550 nm before optimization. (d) Reflectance spectrum and (e) phase spectrum of the phase-change metasurface after optimization (amorphous and crystalline states). (f) Reflectance and phase modulation as functions of crystallization ratio at 1550 nm after optimization.

To verify the effectiveness of the proposed metasurface design concept in enhancing the spatial light field modulation efficiency of phase-change metasurfaces, a gradient-phase metasurface was designed with +1st order diffraction beam deflection to compare the changes in modulation efficiency before and after optimization. The designed gradient-phase metasurface divides the phase into three orders: 0°, 120°, and 240°, as illustrated in Fig. 4(a) and 4(d). Each period consists of 18 metasurface units arranged in groups of 6 to achieve a uniform distribution of phase differences, as illustrated in Fig. 4(b) and 4(e). The polarization direction of the incident light is aligned with the phase gradient direction. According to generalized Snell's law, this gradient-phase metasurface achieves a beam deflection of 4.29° in the direction of the phase gradient at the target wavelength of 1550 nm. The expression of generalized Snell's law is as follows:

$$sin\theta_r - sin\theta_i = \frac{\lambda_0}{2\pi n_i} \times \frac{d\varphi}{dy} \qquad (1)$$

where $\theta_i$ is the angle of incidence, $\theta_r$ is the angle of deflection, $\lambda_0$ is the target wavelength, $d\varphi/dy$ is the phase gradient, and $n_i$ is the refractive index of the incident medium.

To demonstrate the efficiency of the spatial light field phase modulation, systematic numerical simulations were conducted of the reflectance distribution at different wavelengths and diffraction orders of the gradient-phase metasurface before and after optimization, as shown in Fig. 4(c) and 4(f). The results show that the diffracted energy before optimization is primarily concentrated in the 0th order diffraction, which indicates weak modulation of the spatial light field and difficulty in effectively transmitting the optical energy to the desired direction. This phenomenon is primarily due to the fact that the reflectance of metasurfaces before optimization approaches 100% at crystallization ratio $m_1$, while it approaches 0 at ratios $m_2$ and $m_3$. This strong coupling between amplitude and phase modulation implies that a majority of light energy is absorbed, causing the spatial distribution of the light field to be highly uneven, leading to distortion of the reflected light wavefront and reducing the overall efficiency of light field regulation in the device. These findings reveal substantial limitations in the phase state selection and energy conversion efficiency of the metasurface before optimization. After optimization, the distribution of diffracted energy is significantly improved, primarily concentrated in the 0th and +1st orders of diffraction, indicating that the optimized gradient-phase metasurface can effectively guide the energy of incident light to the target diffraction direction. The key to this change is that the reflectance of the optimized metasurface at crystallization ratios $m_1$, $m_2$, and $m_3$ exceeds 46.5%, with minor variations. This indicates that a majority of incident light can be effectively reflected, while amplitude fluctuations exert a lesser influence on phase modulation, thus facilitating efficient modulation of the spatial light field and enabling more flexible transmission of light energy. To further evaluate the optimization effect, a comparison was made of the +1st order and total reflection spectrum before and after the optimization of the gradient-phase metasurface at the target wavelength of 1550 nm, as shown in Fig. 4(g) and 4(h). After optimization, the +1st order diffraction efficiency of the gradient-phase metasurface at this wavelength increased significantly from 13.5% to 51.3%. This significant improvement not only demonstrates the effectiveness of the design optimization but also highlights its substantial potential for achieving efficient spatial light field modulation at specific wavelengths. In addition, a further comparison was made of the +1st order reflectance before and after optimization at different deflection angles by changing the period of the gradient-phase metasurface, as shown in Fig. 4(i). The optical responses and settings of the remaining gradient-phase metasurfaces are detailed in Supplementary Materials S5 to S12. The analysis results indicate that the +1st order diffraction efficiency of the optimized gradient-phase metasurface exceeds 35.6% at deflection angles of 3.68°, 4.29°, 5.16°, 6.45°, and 8.61°. This result not only demonstrates the exceptional optical performance and control capabilities of the gradient-phase metasurface under varying diffraction conditions but also illustrates the broad application potential of the proposed metasurface design concept in spatial light field modulation.

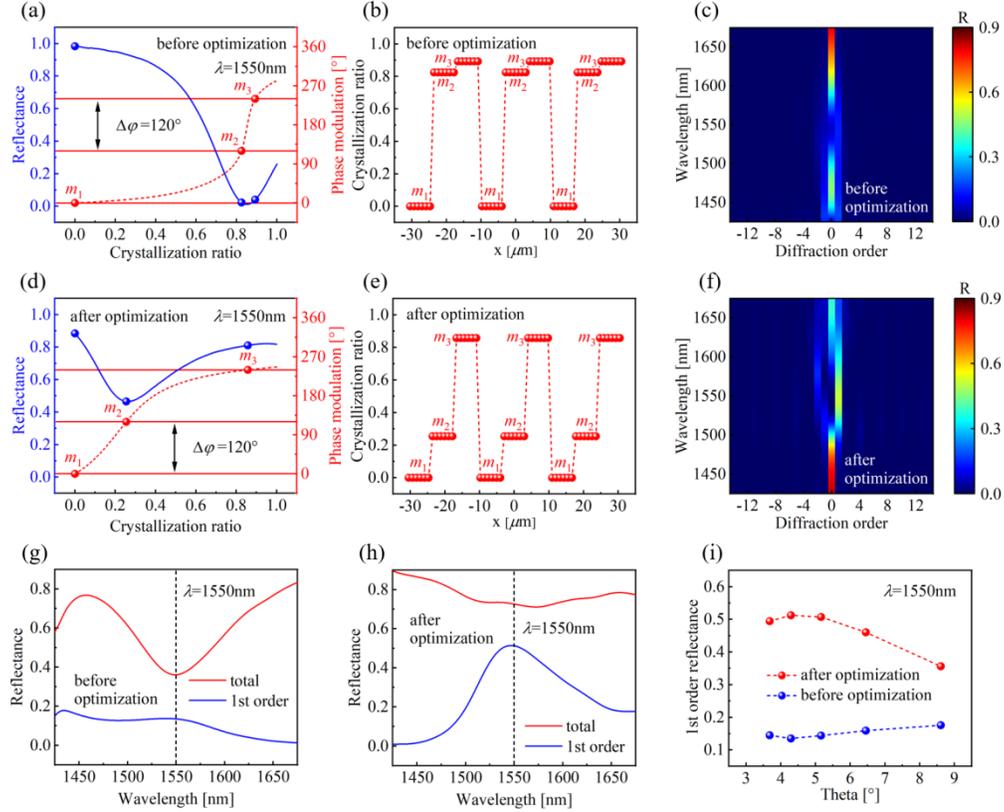

Fig. 4. Comparison of the performance of the gradient-phase metasurface before and after optimization. (a) Phase gradient of the metasurface before optimization. (b) Crystallization ratio distribution along the phase gradient direction before optimization; (c) Reflectance of diffraction orders at various wavelengths before optimization. (d) Phase gradient of the metasurface after optimization. (e) Crystallization ratio distribution along the phase gradient direction after optimization. (f) Reflectance of diffraction orders at various wavelengths after optimization. (g) +1st order and total reflectance spectra before optimization. (h) +1st order and total reflectance spectra after optimization. (i) +1st order reflectance at various deflection angles at a wavelength of 1550 nm before and after optimization.

The current design concept focuses on adjusting the phase state of the resonant wavelength as it transitions through the target wavelength. This allows the resonant mode to be localized in a state of low crystallization ratio, achieving a phase-dominated modulation effect. Similarly, by optimizing the structural parameters, the resonance mode can be localized in a state of high crystallization rati, resulting in an amplitude-dominated modulation effect. Therefore, we replaced the GSST cylinder with an elliptical cylinder to induce anisotropy in the phase-change microstructure and introduce both amplitude-dominated and phase-dominated modulation effects in different polarization directions. In the optimized metasurface, the GSST elliptical cylinder features a short-axis radius of 140 nm and a long-axis radius of 390 nm. The thicknesses of the GSST elliptical cylinder, Au reflective layer, $Al_2O_3$ isolation layer, and $Al_2O_3$ protective layer are maintained at 100 nm, 100 nm, 50 nm, and 20 nm, respectively, with the unit structure period of the metasurface set at 1150 nm. When the incident light's electric field is parallel to the long axis (polarized in the x-direction), the resonance center wavelength of the supported dielectric resonance mode in the phase-change microstructure shifts from 1507 nm to 1894 nm during the phase transition, as illustrated in Fig. 5 (a) and 5(b). At the target wavelength of 1550 nm, a phase-dominant modulation effect is achieved, resulting in a minimum reflectance of 50.9% and a phase modulation depth of 241.0°, as shown in Fig. 5(c).

In contrast, when the electric field direction of the incident light is parallel to the short axis (polarized in the y-direction), the resonance center wavelength of the dielectric resonance mode supported by the phase-change microstructure shifts from 1378 nm to 1540 nm, as shown in Fig. 5(d) and 5(e). At the target wavelength of 1550 nm, a amplitude-dominant modulation effect is observed, with an amplitude modulation depth of 96.4%, while the phase modulation depth is only 79.2°, as shown in Fig. 5(f). This phenomenon demonstrates that incident light polarized in orthogonal directions can produce different light field modulation effects. By imparting anisotropy to the phase-change microstructure based on the existing design concept of metasurfaces, different optical parameters can be effectively modulated in various polarization directions, thereby broadening the scope of application for the current design. Additionally, under specific polarization conditions, a trade-off exists between amplitude and phase modulation of the metasurface. This indicates that the mutual influence of different optical effects must be comprehensively considered during the design process to achieve optimal modulation outcomes.

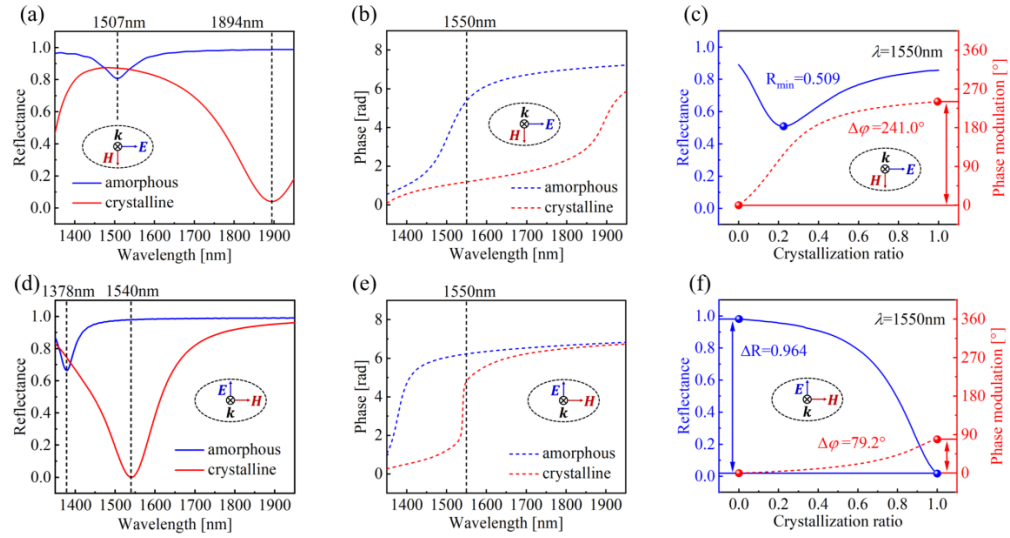

Fig. 5. Analysis of the optical properties of the metasurface for different polarization directions. (a) Reflectance and (b) Phase spectra (amorphous and crystalline) with the electric field parallel to the long axis. (c) Reflectance and phase modulation at a wavelength of 1550 nm as a function of crystallization ratio with the electric field parallel to the long axis. (d) Reflectance and (e) Phase spectra (amorphous and crystalline) with the electric field parallel to the short axis. (f) Reflectance and phase modulation at a wavelength of 1550 nm as a function of crystallization ratio with the electric field parallel to the short axis.

By exploiting the polarization-dependent phase response, efficient dynamic birefringence modulation is achieved, as illustrated in Fig. 6(a). At the target wavelength of 1550 nm, a birefringence modulation depth of approximately 197.6° is achieved during the phase transition of the phase-change microstructure, thereby providing the necessary conditions for dynamic polarization modulation. To detail the change in the polarization state of reflected light, the Stokes parameters (normalized by $S_0$) of the phase-change metasurface during the phase transition are analyzed for incident light with a polarization direction of 45°, as illustrated in Fig. 6(b). At lower crystallization ratios, the incident light exhibits high reflectance in both the x- and y-polarized directions, resulting in $S_1/S_0$ remaining essentially 0 without significant change. In this process, the polarization state of the reflected light is primarily determined by birefringence modulation. Simultaneously, $S_2/S_0$ approaches -1 at a crystallization ratio of 19.5%, while $S_3/S_0$ approaches -1 and 1 at crystallization ratios of 1% and 53%, respectively. As the crystallization ratio increases, the component of incident light in the y-polarization direction is significantly absorbed by the phase-change microstructure, leading to increased energy loss at high crystallization ratios, while the effect of birefringence modulation gradually weakens. This results in $S_1/S_0$ approaching 1 when the crystallization ratio reaches 100%, while $S_2/S_0$ and $S_3/S_0$ both approach 0. These results indicate that the metasurface can dynamically convert reflected light among four polarization states: left-handed circularly polarized light ($|l\rangle$), -45° linear polarization ($|b\rangle$), right-handed circular polarization ($|r\rangle$), and x-direction linear polarization ($|x\rangle$), corresponding to crystallization ratios of 1%, 19.5%, 53%, and 100%, respectively. To intuitively demonstrate the dynamic polarization modulation capability of the metasurface, a polarization Poincaré sphere based on the Stokes parameter coordinate system was employed to analyze dynamic polarization conversion, as illustrated in Fig. 6(c). During crystallization, the polarization state of the reflected light traces a semicircle in the $S_3/S_0$ and $S_2/S_0$ planes, indicating that it is predominantly influenced by the $S_2/S_0$ and $S_3/S_0$ components at this stage, which suggests that birefringence modulation primarily governs the polarization state. Following the increasing absorption of the incident light component in the y-polarization direction by the phase-change microstructure, the polarization state describes a quarter-circle trace in the $S_3/S_0$ and $S_1/S_0$ planes. This indicates that changes in the polarization state are primarily governed by the $S_3/S_0$ and $S_1/S_0$ components, reflecting the combined effects of energy loss in the y-polarization direction and birefringence modulation. This dynamic change process illustrates the precise control capabilities of the polarization state across different phase states, demonstrating that metasurfaces can achieve a broad range of dynamic polarization modulation performance. Simultaneously, the polarization degree of the four polarization states was calculated, yielding a result greater than 94.9%, as depicted in Fig. 6(d). This high polarization degree not only demonstrates the precise control capability of the metasurface over the polarization state but also underscores its potential advantages in optical applications. Additionally, the variations of the four polarization states and the total polarization conversion efficiency were further analyzed, as shown in Fig. 6(e). At the target wavelength of 1550 nm, the total polarization conversion efficiency of the metasurface remains above 43%, with the conversion efficiencies of the four polarization states being 92.336%, 71.973%, 78.76%, and 42.838%, respectively. These results illustrate the effectiveness and flexibility of the metasurface in dynamic polarization modulation. It is important to note that under high crystallization ratios, the total polarization conversion efficiency decreases due to significant energy loss resulting from the absorption of incident light with a y-polarized component by the phase-change microstructure. Although the current design effectively utilizes the low absorption characteristics of the amorphous state, the influence of intrinsic absorption at high crystallization ratios on conversion efficiency remains unavoidable. This finding underscores the necessity of thoroughly considering the absorption characteristics of materials when designing high-efficiency metasurfaces to achieve enhanced polarization modulation performance.

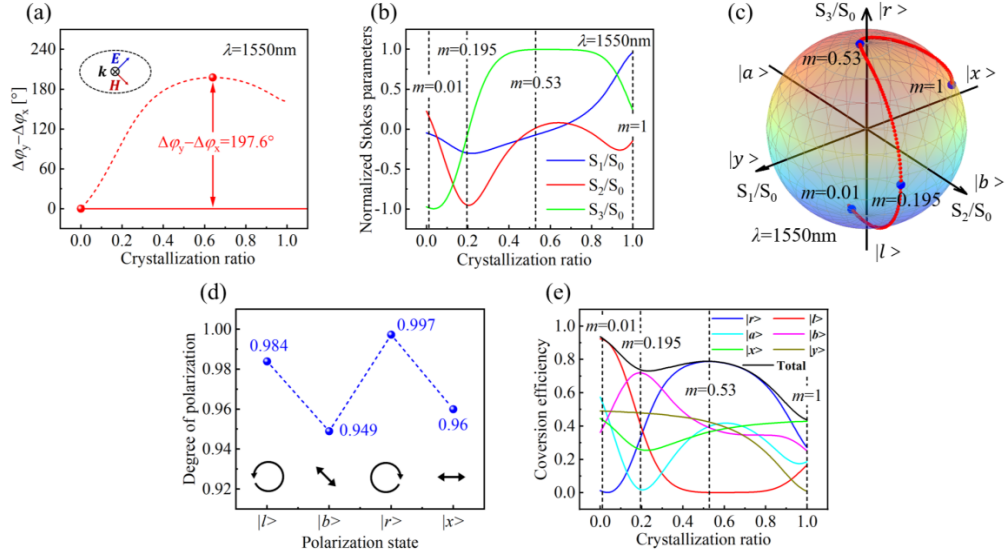

Fig. 6. Analysis of the performance of dynamic polarization modulation. (a) Birefringence modulation as a function of crystallization ratio at the target wavelength of 1550 nm. (b) Change in the Stokes parameters of reflected light during phase transition at the target wavelength of 1550 nm (normalized by $S_0$). (c) The trace of the polarization state of the reflected light on the polarization Poincaré sphere during crystallization at the target wavelength of 1550 nm. (d) The polarization degree of the four polarization states. (e) The variation in the conversion efficiency of the four polarization states.

## 4. Conclusion

In this work, we successfully addresses the efficiency challenges of phase-change metasurfaces in their crystalline state, which has been a significant barrier to their practical application in continuous spatial light modulation. Our metasurface design strategically localizes the phase modulation process in a low crystallization ratio state, significantly reducing the impact of material absorption and enhancing modulation efficiency. The optimized GSST-based metasurface achieves a minimum reflectance of 46.5% and a phase modulation depth of 246.6° at 1550 nm, which are substantial improvements over existing technologies. Furthermore, by introducing structural anisotropy, our design enables independent control of amplitude-dominant and phase-dominant modulation in orthogonal polarization directions. This innovation allows for dynamic polarization modulation, demonstrating the capability for multistate polarization conversion with high efficiency. The ability to achieve multistate polarization conversion is a significant advancement, as it expands the functionality of phase-change metasurfaces beyond simple switching, enabling complex light control applications such as dynamic holography, high-dimensional data storage, and advanced optical communication systems. This work represents a significant step forward in the development of phase-change metasurfaces, paving the way for their widespread integration into advanced photonic systems that demand efficient, continuous, and multifunctional light control.

**Acknowledgment.**

This work was supported by the Joint Funds of the National Natural Science Foundation of China (Grant No. U23A20372) and National Natural Science Foundation of China (NSFC) (62375291).

**Disclosures.** The authors declare no conflicts of interest.

# APPROACHING HIGH-EFFICIENCY SPATIAL LIGHT MODULATION WITH LOSSY PHASE-CHANGE MATERIAL: SUPPLEMENTAL DOCUMENT

## 1. The finite element method

The optical properties of the metasurface are simulated using the finite element method with the commercial software Lumerical FDTD. In the simulation model, periodic boundary conditions are applied to the x and y directions of the unit cell to accurately reflect the periodic structure of the actual metasurface. This configuration ensures that the simulation results accurately represent the optical behavior of the metasurface. Furthermore, to effectively simulate an infinite isotropic medium environment, perfectly matched layers (PMLs) are implemented at the top and bottom of the metasurface unit, significantly reducing the influence of reflected waves on the simulation results. The multipole decomposition of the metasurface is conducted using the commercial software COMSOL Multiphysics. The multipole decomposition algorithm is grounded in relevant literature, providing a solid theoretical foundation for this research and ensuring the accuracy and validity of the results [1-3].

The complex refractive index data for Si, Au, and $Al_2O_3$ are obtained from the material database of the commercial software Lumerical FDTD. The complex refractive index of $Ge_2Sb_2Se_4Te_1$ in both crystalline and amorphous states is sourced from relevant literature [4]. Furthermore, the complex refractive index of $Ge_2Sb_2Se_4Te_1$ in any intermediate state is calculated using the Lorentz-Lorenz equation, which characterizes the relationship between a material's optical properties and its microstructure. The Lorentz-Lorenz equation can be expressed as follows:

$$\frac{\epsilon_{eff}(\lambda) - 1}{\epsilon_{eff}(\lambda) + 2} = m \times \frac{\epsilon_c(\lambda) - 1}{\epsilon_c(\lambda) + 2} + (1 - m) \times \frac{\epsilon_a(\lambda) - 1}{\epsilon_a(\lambda) + 2} \tag{S1}$$

In this equation, $\epsilon_c(\lambda)$ and $\epsilon_a(\lambda)$ represent the complex permittivity of the crystalline and amorphous states at the wavelength $\lambda$, respectively. Here, $m$ denotes the crystallization ratio, while $\epsilon_{eff}(\lambda)$ indicates the effective complex permittivity at crystallization ratio $m$.

## 2. The characterization of resonant modes

At the target wavelength of 1550 nm, both the optimized and unoptimized phase-change metasurfaces exhibit distinct resonance peaks at crystallization ratios of 21% and 85%, respectively. These resonance peaks indicate that the metasurface exhibits a strong response to incident light at the corresponding crystallization ratios (see Fig. S1(a) and S2(a)). To gain further insight into the physical nature of these resonance peaks, we conducted a multipole decomposition of the resonance modes (see Fig. S1(b) and S2(b)). The results indicate that at a wavelength of 1550 nm, the electric dipole (ED) serves as the dominant resonance mode, while the magnetic dipole (MD), toroidal dipole (TD), electric quadrupole (EQ), and magnetic quadrupole (MQ) contribute relatively little to the scattering cross section. This suggests that the electric dipole plays a crucial role in overall light scattering. Moreover, Fig. S1(c) and S2(c) illustrate the real and imaginary parts of the complex amplitude of the reflected light field at 1550 nm for the metasurface before and after optimization, during the phase transition of the GSST cylinder at varying crystallization ratios. This relationship reveals the optical response of the metasurface at different crystallization ratios, indicating that the resonance mode remains in an overcoupled state, effectively modulating the phase of the reflected light field.

To further characterize the excited dielectric resonance modes, Fig. S1(d) to S1(k) and S2(d) to S2(k) display the absolute value distributions of the electric and magnetic fields at the zy, zx, and yx cross-sections at a wavelength of 1550 nm for the phase-change metasurface, both before optimization (at a crystallization ratio of 21%) and after optimization (at a crystallization ratio of 85%). Additionally, the absolute value distributions of the z-component of the electric field and the y-component of the magnetic field on the zy cross-section are presented. The results reveal that the absolute value distributions of the electromagnetic field in these three cross-sections demonstrate a typical dielectric resonance mode dominated by an ED. This further confirms that the ED-dominated dielectric resonance mode is effectively excited. Notably, there is almost no z-component of the electric field on the surface of the Au reflective layer in the zy cross-section. This observation indicates that, despite the presence of the Au reflective layer as a bottom material, it has minimal impact on exciting the plasmon resonance mode. This suggests that the Au reflective layer does not significantly interfere with the excitation of electric dipoles, thereby enhancing the metasurface's ability to control electromagnetic waves within the target spectral range. Consequently, this structural design not only effectively utilizes the dielectric resonance mode but also mitigates energy loss resulting from metal absorption.

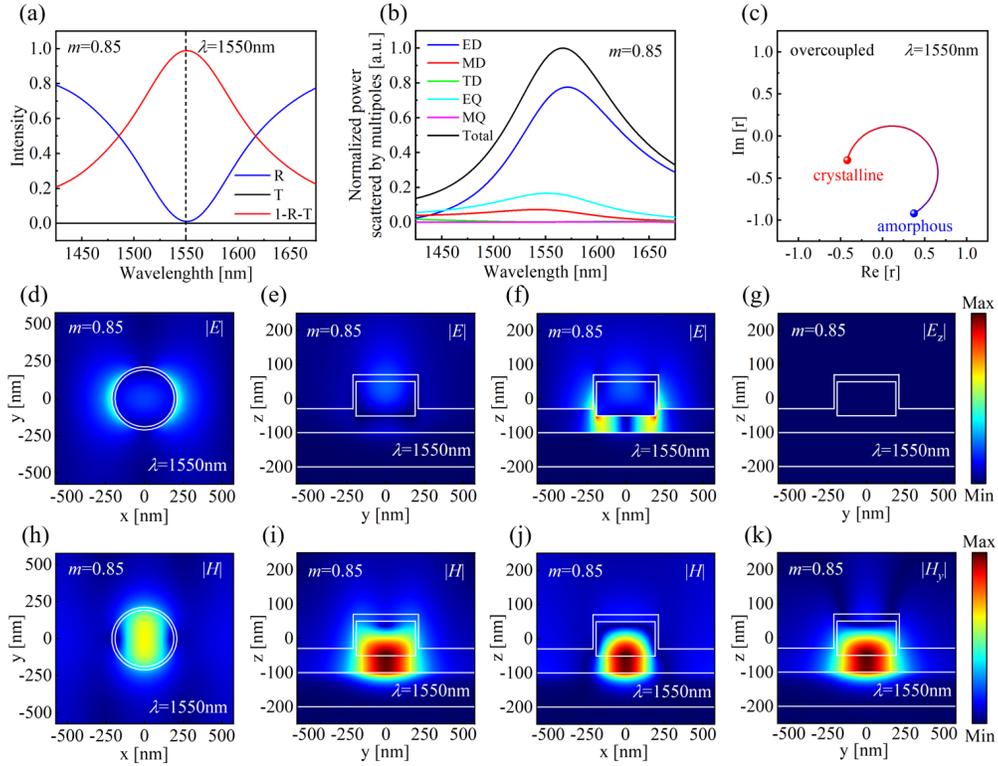

Fig. S1. Analysis of the optical properties of the phase-change metasurface before optimization. (a) Reflectance and transmission spectra of the GSST cylinder with a crystallization ratio of 85%. (b) Multipole decomposition at various wavelengths for the GSST cylinder with a crystallization ratio of 85%. (c) Real and imaginary parts of the complex amplitude of the reflected light field as a function of the crystallization ratio at a wavelength of 1550 nm. Spatial distribution of the absolute value of the electric field at (d) the yx cross-section, (e) the zy cross-section, and (f) the zx cross-section at a wavelength of 1550 nm for the GSST cylinder with a crystallization ratio of 85%. (g) Spatial distribution of the absolute value of the z-component of the electric field in the zy cross-section at a wavelength of 1550 nm for the GSST cylinder with a crystallization ratio of 85%. Spatial distribution of the absolute value of the magnetic field in the (h) the yx cross-section, (i) the zy cross-section, and (j) the zx cross-section at a wavelength of 1550 nm for the GSST cylinder with a crystallization ratio of 85%. (k) Spatial distribution of the absolute value of the y-component of the magnetic field in the zy cross-section at a wavelength of 1550 nm for the GSST cylinder with a crystallization ratio of 85%.

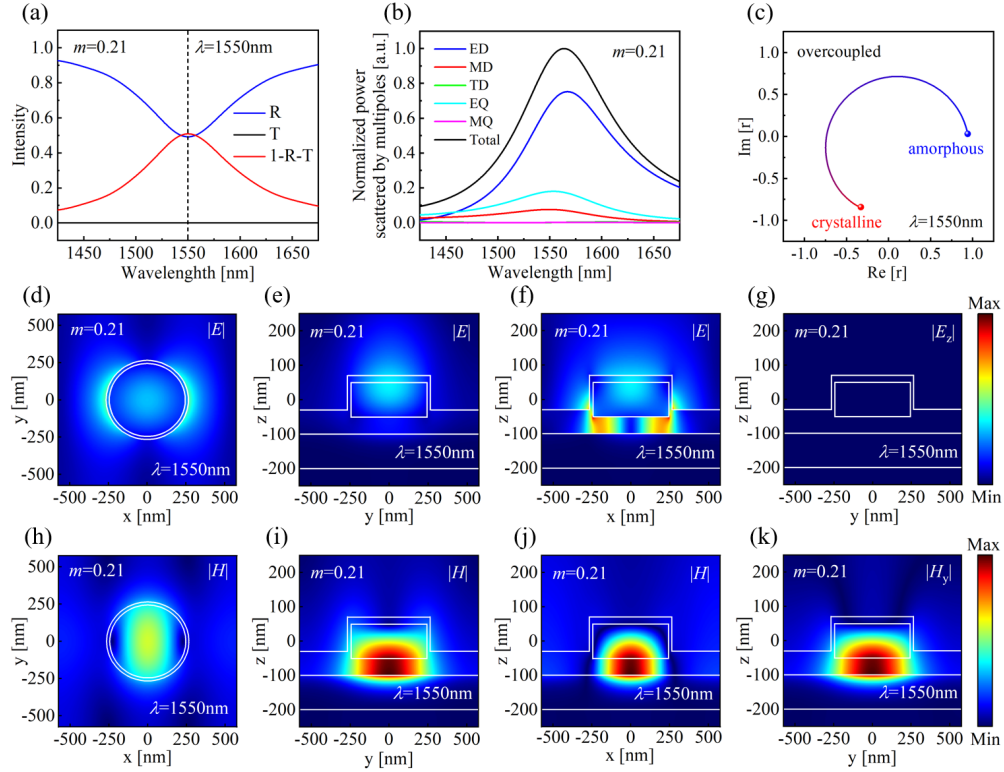

Fig. S2. Analysis of the optical properties of the phase-change metasurface after optimization. (a) Reflectance and transmission spectra of the GSST cylinder with a crystallization ratio of 21%. (b) Multipole decomposition at various wavelengths for the GSST cylinder with a crystallization ratio of 21%. (c) Real and imaginary parts of the complex amplitude of the reflected light field as a function of the crystallization ratio at a wavelength of 1550 nm. Spatial distribution of the absolute value of the electric field at (d) the yx cross-section, (e) the zy cross-section, and (f) the zx cross-section at a wavelength of 1550 nm for the GSST cylinder with a crystallization ratio of 21%. (g) Spatial distribution of the absolute value of the z-component of the electric field in the zy cross-section at a wavelength of 1550 nm for the GSST cylinder with a crystallization ratio of 21%. Spatial distribution of the absolute value of the magnetic field in the (h) the yx cross-section, (i) the zy cross-section, and (j) the zx cross-section at a wavelength of 1550 nm for the GSST cylinder with a crystallization ratio of 21%. (k) Spatial distribution of the absolute value of the y-component of the magnetic field in the zy cross-section at a wavelength of 1550 nm for the GSST cylinder with a crystallization ratio of 21%.

## 3. The principle of the metasurface design concept

In order to further elucidate the proposed design concept of the phase-change metasurface, Fig. S3(a) and S3(b) illustrate the multipole decomposition results of the metasurface before and after optimization under different crystallization ratios at the wavelength of 1550 nm. The analysis indicates that, based on this metasurface design concept, the crystallization ratio at which the peak normalized scattered power of the optimized metasurface is observed has significantly decreased, from the original crystallization ratio of 75% to 11.5%. This significant change not only indicates that the optical properties of the metasurface have been significantly improved, but also demonstrates the capability to achieve strong resonance at a lower crystallization ratio. Additionally, a detailed simulation analysis was conducted on the absorptance of the phase-change metasurface before and after optimization at the target wavelength of 1550 nm for the GSST cylinder and Au reflective layer, as shown in Fig. S3(c). The results indicate that the crystallization ratio at which the peak absorptance of the GSST cylinder at a wavelength of 1550 nm is located has been significantly reduced after the optimization of the metasurface, shifting from the original crystallization ratio of 86% to 30%. This change clearly indicates that the optimization has yielded significant improvements in reducing the impact of the intrinsic absorption of PCMs on the efficiency of spatial light field regulation at high crystallization ratios. Specifically, the peak absorptance of the GSST cylinder has been significantly reduced, from 88.5% to 39.2%. This result not only reflects the greatly improved energy transmission efficiency of the optimized metasurface but also demonstrates its potential application value at low crystallization ratios. In contrast, the peak absorbance of the Au reflective layer remains largely unchanged before and after optimization, and its proportion in the total absorption is consistently minimal. This further underscores that the Au reflective layer is not the primary source of energy absorption and does not significantly influence the optimization effects of the metasurface within the overall system. This indicates that the success of the optimization is primarily dependent on the properties of the GSST material itself and its performance at different crystallization ratios. To further explore this point, the material of the Au reflective layer was replaced with a perfect electric conductor. During the phase transition of the phase-change microstructure, it was observed that the minimum reflectance at the wavelength of 1550 nm was further increased to 58.6%, and a phase modulation depth of 242.0° was achieved, as shown in Fig. S3(d). These results indicate that during the design and fabrication process, replacing the reflective layer material with a dielectric material that provides high reflectance and low absorptance at the target wavelength is feasible, or selecting a distributed Bragg reflector (DBR) composed of two dielectric materials with different refractive indices exhibiting low absorptance characteristics at the target wavelength is also an option. This not only further optimizes the optical performance but also enhances the operability and efficiency of the device. To more intuitively illustrate the effectiveness of the proposed metasurface design concept in mitigating the intrinsic absorption of PCMs on the efficiency of spatial light field modulation, Fig. S4(a) to S4(c) and S5(a) to S5(c) show the spatial distribution of the absorption at 1550 nm on the yx, zy, and zx cross-sections of the phase-change metasurface before optimization with a crystallization ratio of 21% and after optimization with a crystallization ratio of 85%. The results clearly show that the absorptance distribution of the optimized metasurface exhibits substantial variations across different cross-sections. After optimization, the absorptance of the GSST cylinder is significantly reduced, indicating that this design effectively enhances the optical efficiency of the metasurface, enabling effective resonance excitation at lower crystallization ratios. This optimization not only improves the optical responsiveness of the material but also indicates its potential for practical applications, especially in optical devices that require high efficiency. In contrast, the absorption of the Au reflective layer remains largely unchanged, accounting for a relatively minor proportion of the total absorption. This observation reinforces the validity of the proposed design concept for phase-change metasurfaces, especially at high crystallization ratios. It significantly reduces the impact of the intrinsic absorption caused by the PCMs on the efficiency of spatial light field manipulation, offering novel strategies for enhancing the performance of photonic devices.

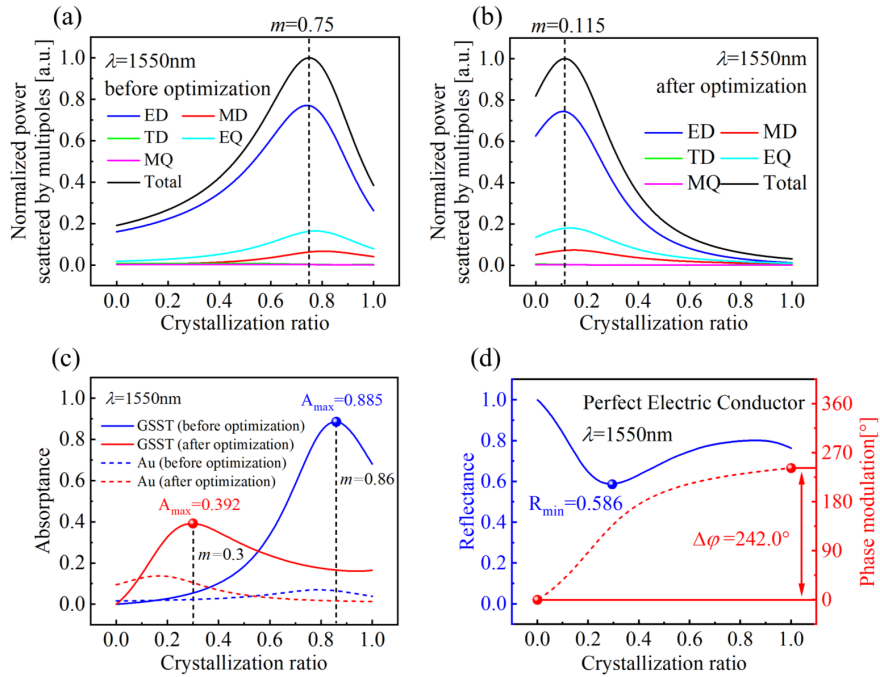

Fig. S3. Analysis of multipole decomposition and absorptance of the phase-change metasurface. (a) Multipole decomposition of the phase-change metasurface before optimization across various crystallization ratios at a wavelength of 1550 nm. (b) Multipole decomposition of the phase-change metasurface following optimization across various crystallization ratios at a wavelength of 1550 nm. (c) Absorptance analysis of the GSST cylinder and Au reflective layer of the phase-change metasurface before and after optimization across various crystallization ratios at a wavelength of 1550 nm. (d) Reflectance and phase modulation of the optimized phase-change metasurface after substituting the Au reflective layer with a perfect electric conductor as a function of the crystallization ratio at a wavelength of 1550 nm.

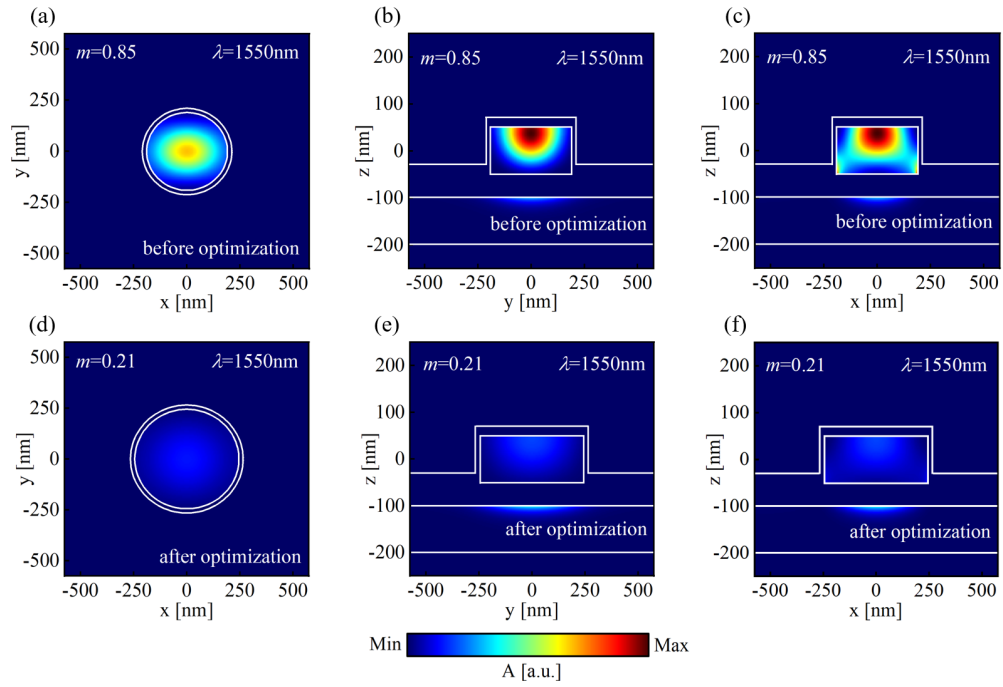

Fig. S4. Analysis of the spatial distribution of absorption for the phase-change metasurface. Spatial distribution of absorption in the phase-change metasurface before optimization at a wavelength of 1550 nm on (a) the yx cross-section, (b) the zy cross-section, and (c) the zx cross-section with a GSST cylinder crystallization ratio of 85%. Spatial distribution of absorption on (d) the yx cross-section, (e) the zy cross-section, and (f) the zx cross-section at a wavelength of 1550 nm for the optimized phase-change metasurface featuring a GSST cylinder crystallization ratio of 21%.

## 4. Setup and optical response of gradient-phase metasurfaces with different deflection angles

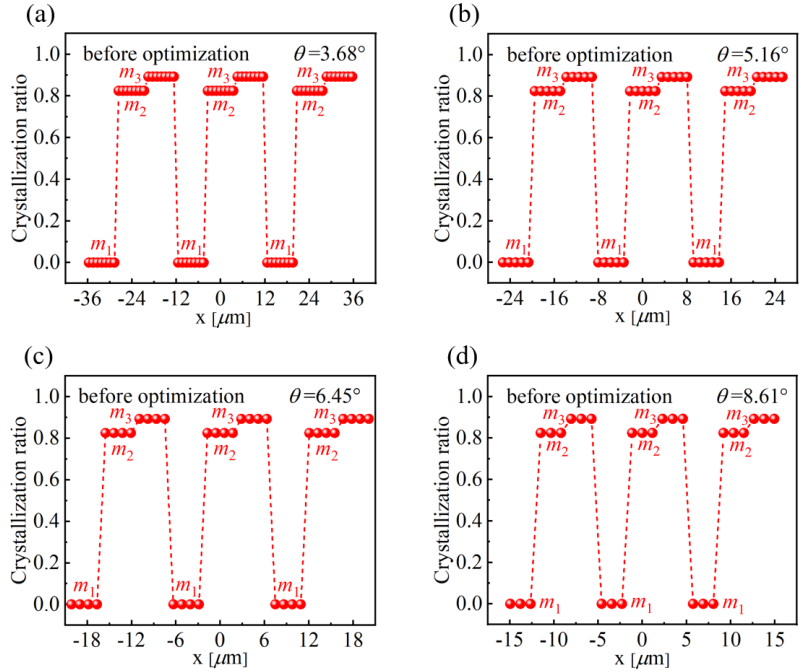

Fig. S5. Crystallization ratio distribution in the direction of the phase gradient for gradient phase metasurfaces with different deflection angles before optimization. (a) Crystallization ratio distribution in the direction of the phase gradient for a deflection angle of 3.68°. (b) Crystallization ratio in the direction of the phase gradient for a deflection angle of 5.16°. (c) Crystallization ratio in the direction of the phase gradient for a deflection angle of 6.45°. (d) Crystallization ratio in the direction of the phase gradient for a deflection angle of 8.61°.

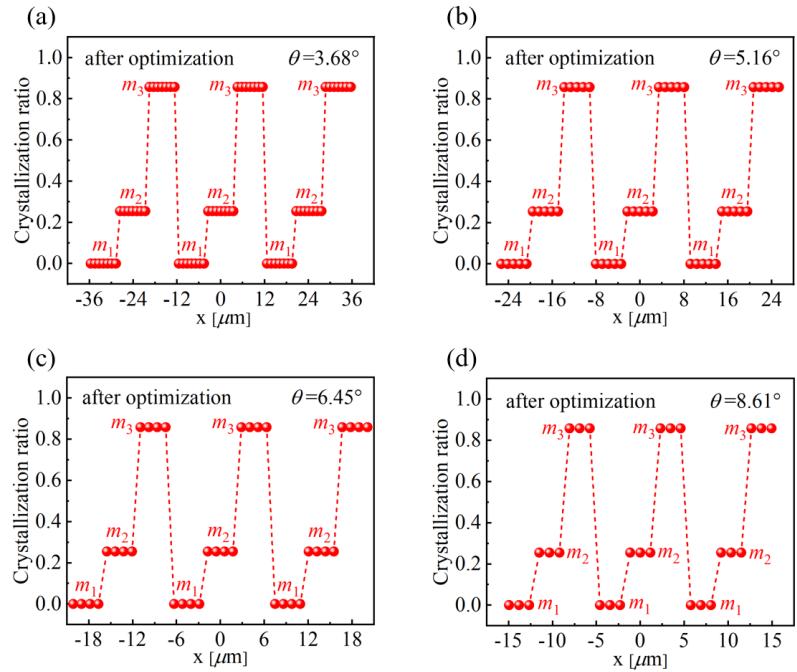

Fig. S6. Crystallization ratio distribution in the direction of the phase gradient for gradient phase metasurfaces with different deflection angles after optimization. (a) Crystallization ratio

distribution in the direction of the phase gradient for a deflection angle of 3.68°. (b) Crystallization ratio in the direction of the phase gradient for a deflection angle of 5.16°. (c) Crystallization ratio in the direction of the phase gradient for a deflection angle of 6.45°. (d) Crystallization ratio in the direction of the phase gradient for a deflection angle of 8.61°.

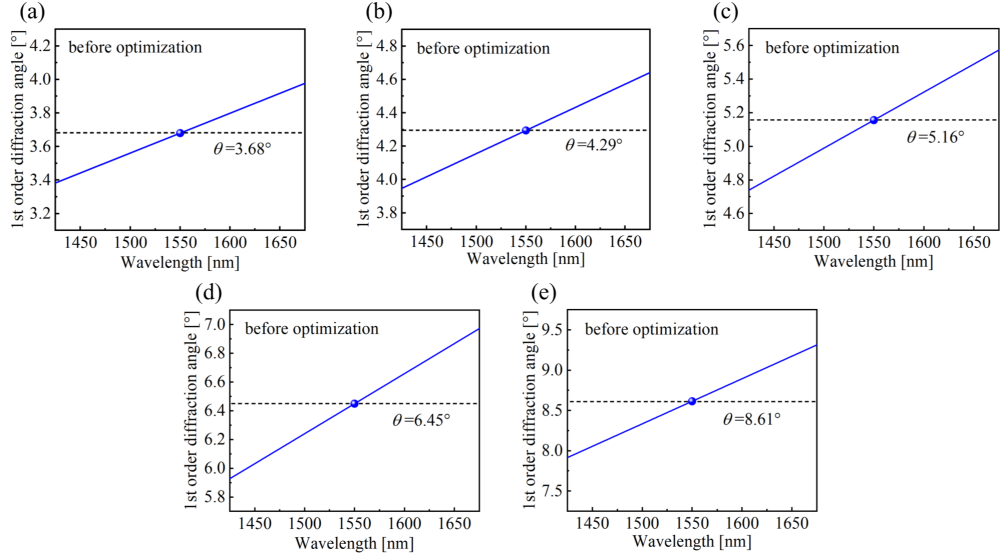

Fig. S7. Analysis of +1st order diffraction angles of gradient phase metasurfaces with different deflection angles before optimization. (a) +1st order diffraction angles of gradient phase metasurfaces with a deflection angle of 3.68° at different wavelengths. (b) +1st order diffraction angles of gradient phase metasurfaces with a deflection angle of 4.29° at different wavelengths. (c) +1st order diffraction angles of the gradient-phase metasurfaces with a deflection angle of 5.16° at different wavelengths. (d) +1st order diffraction angles of the gradient-phase metasurfaces with a deflection angle of 6.45° at different wavelengths. (e) +1st order diffraction angles of the gradient-phase metasurfaces with a deflection angle of 8.61° at different wavelengths.

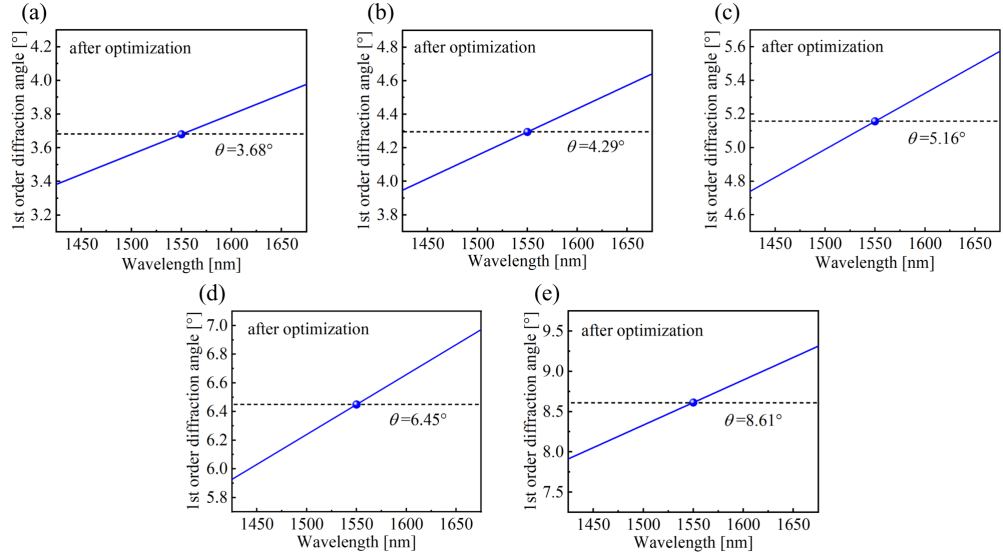

Fig. S8. Analysis of +1st order diffraction angles of gradient phase metasurfaces with different deflection angles after optimization. (a) +1st order diffraction angles of gradient phase metasurfaces with a deflection angle of 3.68° at different wavelengths. (b) +1st order diffraction angles of gradient phase metasurfaces with a deflection angle of 4.29° at different wavelengths.

(c) +1st order diffraction angles of the gradient-phase metasurfaces with a deflection angle of 5.16° at different wavelengths. (d) +1st order diffraction angles of the gradient-phase metasurfaces with a deflection angle of 6.45° at different wavelengths. (e) +1st order diffraction angles of the gradient-phase metasurfaces with a deflection angle of 8.61° at different wavelengths.

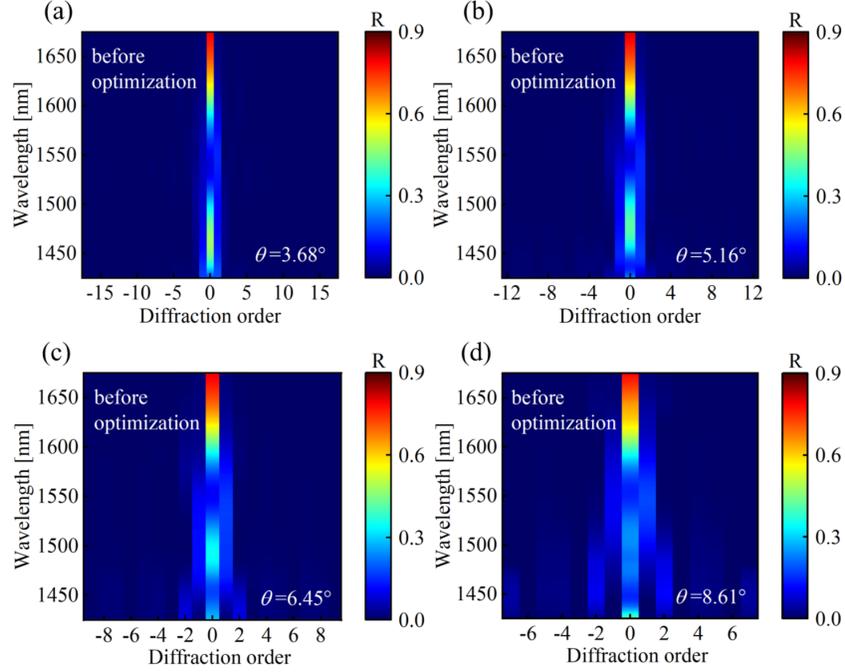

Fig. S9. The reflectance distribution of the gradient phase metasurfaces with different deflection angles at different wavelengths before optimization. (a) The reflectance distribution of the gradient phase metasurfaces with a deflection angle of 3.68° at different wavelengths. (b) The reflectance of the gradient phase metasurfaces with a deflection angle of 5.16° at different wavelengths. (c) The reflectance of gradient-phase metasurfaces with a deflection angle of 6.45° for different diffraction orders at different wavelengths. (d) The reflectance of gradient-phase metasurfaces with a deflection angle of 8.61° for different diffraction orders at different wavelengths.

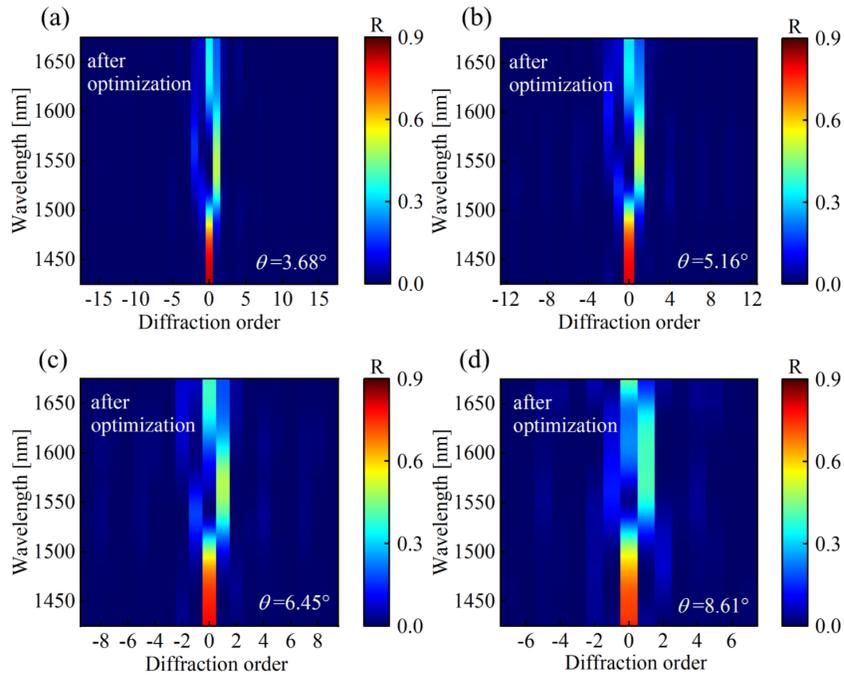

Fig. S10. The reflectance distribution of the gradient phase metasurfaces with different deflection angles at different wavelengths after optimization. (a) The reflectance distribution of the gradient phase metasurfaces with a deflection angle of 3.68° at different wavelengths. (b) The reflectance of the gradient phase metasurfaces with a deflection angle of 5.16° at different wavelengths. (c) The reflectance of gradient-phase metasurfaces with a deflection angle of 6.45° for different diffraction orders at different wavelengths. (d) The reflectance of gradient-phase metasurfaces with a deflection angle of 8.61° for different diffraction orders at different wavelengths.

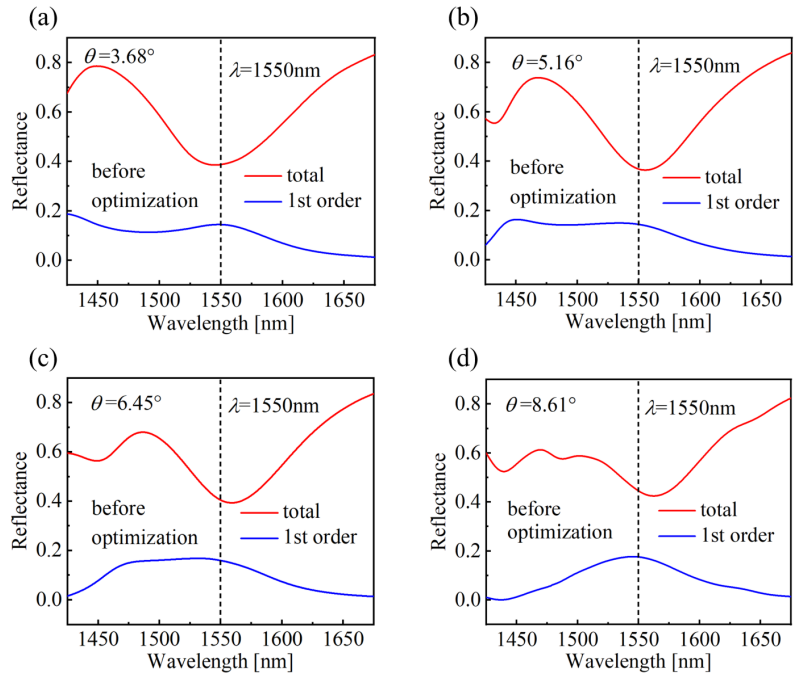

Fig. S11. The +1st-order and total reflectance of the gradient-phase metasurfaces with different deflection angles at different wavelengths before optimization. (a) The +1st-order and total reflectance of the gradient-phase metasurfaces with a deflection angle of 3.68° at different wavelengths. (b) The +1st-order and total reflectance of the gradient-phase metasurfaces with a deflection angle of 5.16° at different wavelengths. (c) The +1st-order and total reflectance of gradient-phase metasurfaces with a deflection angle of 6.45° at different wavelengths. (d) The +1st-order and total reflectance of gradient-phase metasurfaces with a deflection angle of 8.61° at different wavelengths.

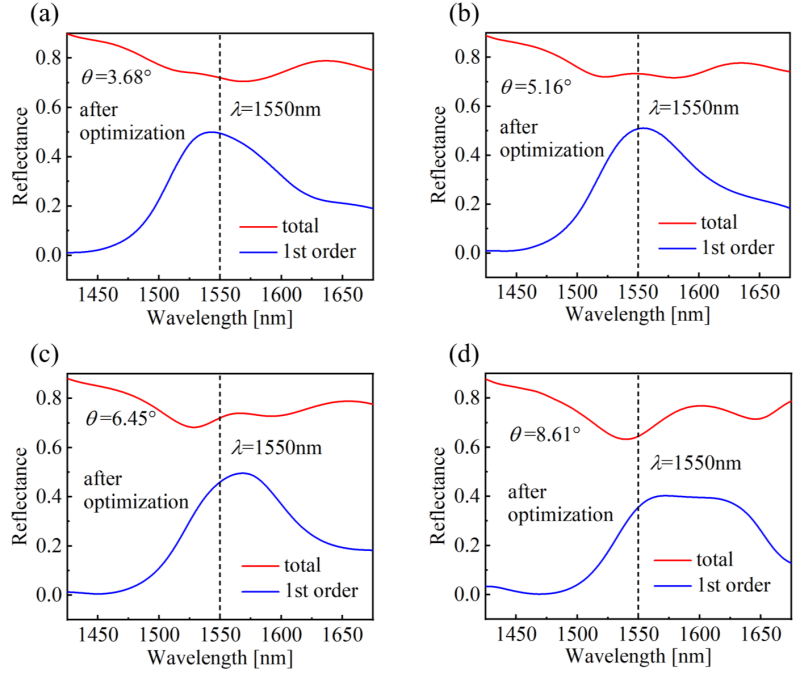

Fig. S12. The +1st-order and total reflectance of the gradient-phase metasurfaces with different deflection angles at different wavelengths after optimization. (a) The +1st-order and total reflectance of the gradient-phase metasurfaces with a deflection angle of 3.68° at different wavelengths. (b) The +1st-order and total reflectance of the gradient-phase metasurfaces with a deflection angle of 5.16° at different wavelengths. (c) The +1st-order and total reflectance of gradient-phase metasurfaces with a deflection angle of 6.45° at different wavelengths. (d) The +1st-order and total reflectance of gradient-phase metasurfaces with a deflection angle of 8.61° at different wavelengths.